\documentclass[aps,prb,twocolumn,groupedaddress,showpacs,showkeys]{revtex4}
\usepackage{amsmath, color}
\usepackage{graphicx}

\begin{document}
\def\c2x2{$c(2\times{}2)$}
\def\feni{Fe$_{64}$Ni$_{36}$}

\title{Chemical ordering and composition fluctuations\\ at the (001) surface of the \feni{} Invar alloy}



\author{M. Ondr\'a\v{c}ek}
\affiliation{Institute of Physics ASCR, Na Slovance 2, CZ-182 21
Praha 8, Czech Republic}

\author{F. M\'aca}\email{maca@fzu.cz}
\affiliation{Institute of Physics ASCR, Na Slovance 2, CZ-182 21
Praha 8, Czech Republic}

\author{J. Kudrnovsk\'y}
\affiliation{Institute of Physics ASCR, Na Slovance 2, CZ-182 21
Praha 8, Czech Republic}

\author{J. Redinger}
\affiliation{Center for Computational Materials Science, Getreidemarkt 9/134,
A-1060 Vienna, Austria}

\author{A. Biedermann}
\affiliation{Institut f\"ur Materialphysik, University of Vienna,
Strudlhofgasse 4, A-1090 Vienna, Austria}

\author{C. Fritscher}\altaffiliation{Present address: Institute of Materials Science and
Technology,
Vienna University of Technology, Favoritenstrasse 9-11, 1040
Vienna, Austria}
 \affiliation{Institut f\"ur
Allgemeine Physik, Vienna University of Technology, Wiedner Hauptstr. 8-10,
A-1040 Vienna, Austria}

\author{M. Schmid}
 \affiliation{Institut f\"ur
Allgemeine Physik, Vienna University of Technology, Wiedner Hauptstr. 8-10,
A-1040 Vienna, Austria}

\author{P. Varga}
\affiliation{Institut f\"ur Allgemeine Physik, Vienna University of
Technology, Wiedner Hauptstr. 8-10, A-1040 Vienna, Austria}

\date{November 13, 2006}

\begin{abstract}

We report on a study of (001) oriented fcc Fe-Ni alloy surfaces
which combines first-principles calculations and low-temperature
STM experiments. Density functional theory calculations show that
Fe-Ni alloy surfaces are buckled with the Fe atoms slightly
shifted outwards and the Ni atoms inwards. This is consistent with
the observation that the atoms in the surface layer can be
chemically distinguished in the STM image: brighter spots
(corrugation maxima with increased apparent height) indicate iron
atoms, darker ones nickel atoms. This chemical contrast reveals a
\c2x2{} chemical order (50\%~Fe) with frequent Fe-rich defects on
the \feni{}(001) surface. The calculations also indicate that
subsurface composition fluctuations may additionally modulate the
apparent height of the surface atoms. The STM images show that
this effect is pronounced compared to the surfaces of other
disordered alloys, which suggests that some chemical order and
corresponding concentration fluctuations exist also in the
subsurface layers of Invar alloy. 
In addition, detailed electronic structure calculations allow us
to identify the nature of a distinct peak below the Fermi level
observed in the tunneling spectra. This peak corresponds to a
surface resonance band which is particularly pronounced in
iron-rich surface regions and provides a second type of chemical
contrast with less spatial resolution but one that is essentially
independent of the subsurface composition.
\end{abstract}


\maketitle

\section{Introduction}

The properties of alloy surfaces are of great importance for
surface- and interface-related phenomena like segregation and
catalysis, furthermore, examination of the surface makes it
possible to study bulk properties not easily accessible by other
methods. Scanning tunneling microscopy (STM) and spectroscopy
(STS) have become indispensable methods for imaging alloy surfaces
\cite{PtNi111,PtRh,schmid02,heben99} and surface alloys
\cite{cham92,nagl94,niel03} with atomic resolution and chemical
contrast. This paper focuses on the interpretation of STM/STS data
of a transition metal alloy, the \feni{} alloy, based on first
principles calculations. In the course of this analysis two
separate modes of chemical contrast in STM images are discussed:
First, apparent height differences of different chemical species
providing chemical contrast in an atom-by-atom fashion; Second,
chemical contrast based on 
specific features of the tunneling spectra, in particular surface
state peaks.

On many transition metal surfaces, surface states and resonances
near the Fermi level have been detected by
STS\cite{cromm93,FeSS,hofer01,Bieder96,CrFe,bisch01}. However,
according to experience and the most basic theory of the STM
\cite{ters85} only a small fraction of surface states,
specifically those which are sufficiently delocalized
perpendicular to the surface, can by seen by the STM, i.e.,
contribute to the tunneling current. A class of surface resonances
accessible by STS is that with $s-p_z-d_{z^2}$ symmetry at the
$\bar{\Gamma }$-point observed on the bcc(001) surfaces of
Fe\cite{FeSS}, Cr\cite{CrFe}, V\cite{bisch01}, W\cite{post80}, and
their surface alloys Cr-Fe\cite{CrFe} and Si-Fe\cite{Bieder96},
and on the hcp(0001) surface of Gd/W(110)\cite{rehb03}. These
states may serve as an extra source of chemical contrast as alloy
components tend to locally quench the surface resonance as
demonstrated for the dilute surface alloy Cr/Fe(001) \cite{CrFe}
and the chemically ordered surface alloy
Si/Fe$_{96.4}$Si$_{3.6}$(001)\cite{Bieder96}.

In this article we apply these ideas to the fcc \emph{alloy} surface
\feni{}(001) which is different from the mentioned \emph{surface} alloys in at
least two ways: (1) Subsurface composition fluctuations\cite{schi99,robe99}
lead to relatively strong variations of the apparent height of the surface
atoms on the nanometer scale, which makes the chemical assignment of the
individual atoms, successfully done for the alloy surfaces
Pt$_{25}$Ni$_{75}$(111)\cite{PtNi111} or Pt$_{50}$Rh$_{50}$(001)\cite{PtRh},
rather difficult. (2) This surface does not show clusters or chains of Fe or
Ni atoms which a priori could be considered sufficiently large to support 1D
or 2D surface resonances characteristic for the elements constituting the
alloy.

\feni{} has a very low thermal expansion for temperatures up to
its Curie temperature of approximately 500 K and is therefore
known under the trademark Invar since its discovery more than 100
years ago. Invar and other iron-rich Fe-Ni alloys are applied in
numerous thermally stabilized or matched components. The thermal
anomalies of these materials are explained by a temperature
dependent atomic volume of Fe in the ferromagnetic state
counteracting ``regular'' thermal expansion due to lattice
anharmonicity (cf., e.g.,
Refs.~\onlinecite{mohn91,entel93,schi99}). A non-collinear
magnetic ordering has been predicted \cite{schi99} and observed
recently \cite{willis05}, which depends on composition
fluctuations and chemical order\cite{schi99,robe99,cris02}.
Therefore, local properties like chemical order, surface and
subsurface composition, and surface electronic structure 
are very relevant to deepen the understanding of this
technologically important system.

After presenting the experimental results, we first investigate the surface
geometry of Fe-Ni(001) alloys from first-principles, in particular, we will
demonstrate that the alloy surface is buckled and this buckling allows us to
explain the chemical contrast observed in atomically resolved STM images.
Second, we present a detailed first-principles investigation of the surface
electronic structure of the \feni{}(001) alloy surface and show that the
surface resonance found just below the Fermi level of Fe-rich surfaces is in
accordance with the local spectroscopic features found in the STM experiment.

\section{Experiment}
\subsection{Setup and sample preparation}

The experiments were done using a low-temperature scanning
tunneling microscope (LT-STM) mounted inside of cryo-shields
cooled by a bath cryostat filled with liquid N$_2$ or liquid He
(Omicron LT-STM). The microscope is operated with
electrochemically etched W tips conditioned by sputtering with
Ar$^+$ under ultrahigh-vacuum (UHV) conditions. The LT-STM system
is equipped with an Auger electron spectroscopy (AES) system for a
surface sensitive chemical analysis. A separate UHV chamber was
used for low energy electron diffraction (LEED) measurements at
300 K and above to probe the temperature dependence of the surface
chemical order. The pressure during measurement in the separate
STM chamber of the LT-STM system is around $1\times{}10^{-11}$
mbar, and in the chamber with the LEED system in the $10^{-11}$
mbar range.

The \feni{}(001) single crystal was typically prepared by sputtering with 2
keV Ar$^+$ ions and subsequent annealing at 770~K for 10~minutes.

\subsection{Chemical contrast and surface order}

\begin{figure}
\begin{center}
\includegraphics[scale=0.5, bb=60 30 500 810, clip]{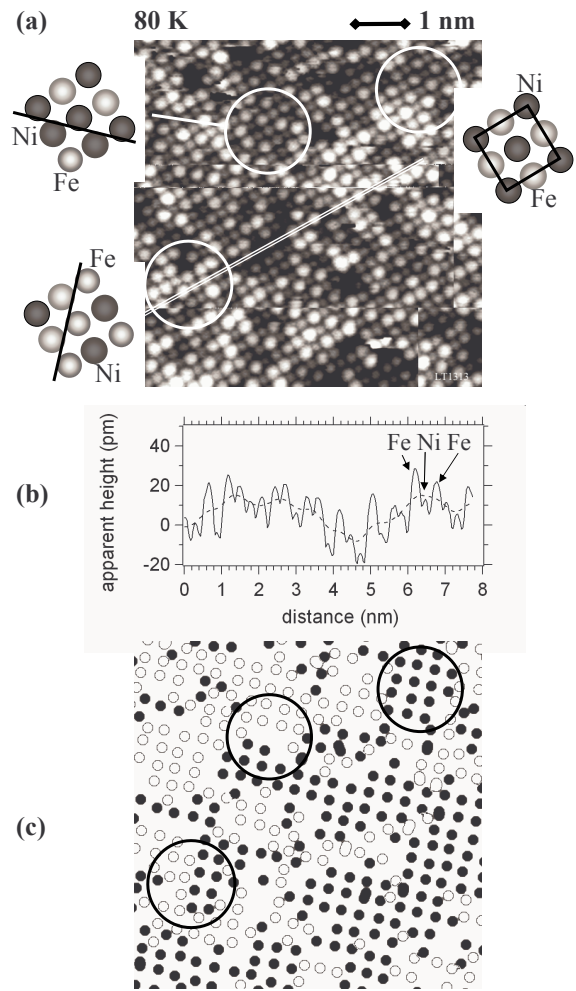}
\end{center}
\caption{(a) STM image of the \feni{}(001) surface (sample voltage
-2~mV, tunneling current 8~nA). The three details highlight a
\c2x2{}-ordered area (unit cell marked by square) and both a short
Ni-rich and a short Fe-rich anti-phase domain boundary segment.
(b) Profile of the STM image along the white line in (a). The
dashed average curve indicates a long-wavelength buckling on the
nanometer scale in contrast to the short-wavelength buckling
distinguishing Fe and Ni atoms. (c) Map of automatically detected
Fe atom locations (corrugation maxima) in the STM image omitting
the Ni atoms, with each of the Fe atoms assigned to one of the two
$(2\times{}2)$ anti-phase sublattices (open and full circles,
occasional oval shapes are due to the detection of more than one
corrugation maximum per atom caused by image errors).}
\end{figure}

Figure 1 shows an STM image of the \feni{}(001) surface at 80 K. The
distinctly different brightness (apparent height) of adjacent atoms indicates
a chemical contrast between Fe and Ni atoms. Images from similar \feni{}
surfaces covered by additionally deposited Fe (about 1 monolayer) in order to
artificially increase the Fe content suggest that the ``dark'' species is Ni.
This atom-by-atom chemical contrast reveals that the surface is locally
ordered with a \c2x2{} superstructure [top right detail in Fig. 1(a)]. Since
the Fe surface concentration is higher than 50\% (cf. next paragraph), the
excess Fe has to be inserted as Fe-rich defects, which often are short pieces
of Fe-rich anti-phase domain boundaries (bottom left detail). Occasionally,
however, also Ni-rich defects are visible in the STM images (middle detail),
indicating that the disorder is larger than required to accommodate the excess
Fe. A LEED experiment probing the temperature dependence of the
(1/2,~1/2)-type superstructure spots normalized to (1,~0) integer spots shows
a reversible rather abrupt disappearance and reappearance of the signal near
550 K indicating an order-disorder transition there. Atomically resolved STM
images with up to $20\times{}20$ nm frame size suggest typical anti-phase
domain diameters between 1 and 10 nm at 80~K.

The chemical analysis by automatically counting bright and dark atoms
based on their apparent height is very unprecise (46\%$\pm$12\% Ni) since
the histogram of apparent atom height values does not show a clear bimodal
distribution of dark and bright atoms. The reason is a significant
buckling of the surface in the entire spatial wavelength range of the
image [dashed line in Fig.~1(b)], interfering with the buckling providing
the chemical contrast between adjacent atoms. Only strong high pass
filtering (not shown in Fig.~1), leaving only height differences of
adjacent atoms in the image, leads to separable peaks in the histogram of
apparent atom heights, which can be fitted by a double gaussian. Combining
the resulting Ni concentration with results from manual atom counting and
a quantitative AES analysis, using the surface sensitive low-energy peaks
of Fe and Ni, results in a Ni concentration of 41\% with an estimated
error of $\pm$2\% for the annealed \feni{} surface---a small Ni enrichment
relative to the nominal bulk concentration of 36\%.

The buckling of the surface layer on the nanometer scale may be
explained by the subsurface distribution of larger Fe and smaller
Ni atoms. Since there exists no long-range order in the \feni{}
bulk, the first hypothesis assumes a fully random order of the
subsurface Fe and Ni atoms, which implies random composition
fluctuations. However, the surfaces of the disordered alloy
Pt$_{25}$Ni$_{75}$(111)\cite{PtNi111}, which shows weak surface
order, or of the alloy Pt$_{25}$Rh$_{75}$(100)\cite{heben99} do
not show a comparable surface buckling on the nanometer scale.
Therefore most likely the second hypothesis applies which assumes
subsurface local order as observed in the \feni{} bulk by element
specific x-ray scattering experiments\cite{robe99}. This would be
an arrangement of subsurface Fe and Ni atoms which is rather
similar to that seen on the surface with \c2x2{} ordered regions
(50\% Fe) interrupted by frequent small Fe-rich domain boundary
fragments. The corresponding composition fluctuations would then
have a characteristic length scale above 1~nm, i.e., the diameter
of Fe-rich defects or ordered domains, as observed, while
fluctuations in randomly distributed alloys are strong only at
very local scales, which have less influence on the buckling of
adjacent atomic layers. We remark, however, that it is likely that
the chemical order on the surface is higher than that in the
subsurface layers due to smaller barriers for atom exchange on the
surface and the low order-disorder transition temperature of the
system around 550~K, i.e., a temperature which seems prohibitive
for efficient bulk diffusion (assuming equal transition
temperatures for the surface and the bulk).

As mentioned at the beginning of this section, this convolution of surface and
subsurface effects make the chemical analysis of the surface based on apparent
atom heights rather cumbersome. In addition, in many cases the STM images do
not show the individual metal atoms and the brightness variations cannot be
unambiguously assigned to either surface chemical contrast or surface
buckling. However, tunneling spectroscopy may be used to separate surface and
subsurface effects -- although with less spatial resolution and less precision
as demonstrated in the next section.

\subsection{Chemical contrast by scanning tunneling spectroscopy}

\begin{figure}[hbt]
\begin{center} \includegraphics[scale=0.45, bb=30 250 560 820, clip]{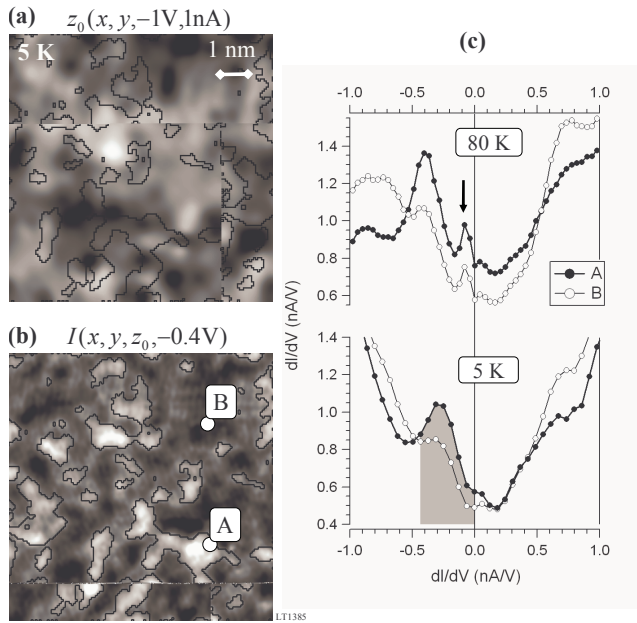}
\end{center}
\caption{(a) Constant current image of the \feni{}(001) surface at 5~K, 1~nA
tunneling current, and -1~V sample voltage. (b) Corresponding current map at
-0.4~V, showing the intensity of a surface resonance below the Fermi-level
(shaded area in 5 K spectrum to the right). (c) dI/dV(V) spectra at 5~K
corresponding to the images at left and of a separately prepared sample at
80~K. Shown are spectra near extremal points of the surface resonance
intensity, both for maximum (A) and minimum (B) intensity. The small peak at
-0.1~V in the 80~K spectra is a tip-related state
 (arrow). The contour lines in (a) and (b) encompass the relatively
small surface fraction that shows spectra which are more similar
(smaller root mean square deviation) to spectrum A than to
spectrum B. The remaining larger surface fraction shows spectra
which are more similar to spectrum B. Evidently, the surface
fraction with significant surface resonance intensity is rather
small (26\% in this data set) consistent with the assumption that
it marks Fe-rich regions.}
\end{figure}

Figure 2 shows STS images and tunneling spectra of the \feni{}(001) surface at
5~K and spectra only for a surface at 80 K. Data acquisition was done by
measuring I/V curves for every image point with the tip position $z_0$
adjusted to 1~nA at -1~V tunneling voltage in every image point [Fig.~2(a)],
followed by numerical differentiation. The spectra shown in Fig.~2(c) are
averages over a few tens of individual spectra from equivalent image points of
the same STS image without additional smoothing. Both experiments, 5 and 80 K,
show a distinct peak at 0.35$\pm$0.1~eV below the Fermi edge, which is
intense, however, only in a very small fraction of the surface. The first
principles calculations in the following sections will describe the character
of the underlying electronic states in detail. In this section we will use the
term ``surface resonance peak'' for the sake of simplicity. The small peak
shift of 0.1 eV between 5 and 80 K is not significant within the precision of
this measurement. The small peak visible in the 80 K measurement [arrow in
Fig.~2(a)], appears in the spectrum of every image point in this data set
only, which is characteristic for a tip-related state.

The peak shape is rather symmetric, indicating little dispersion
of that part of the surface resonance band that is visible for the
STM (cf. Sect.~\ref{surfelstruc}). The width (FWHM) is about 250
meV, which is quite large compared to similar peaks on the
Cr(001)\cite{hank05} or Gd(0001)\cite{rehb03} surface at 5 K.
However, it is comparable to the width of the resonance of single
Fe atoms on the Pt(111) surface at 5 K ($\approx{}$300
meV)\cite{crom93b}. The calculations in the next sections show
that the peak can indeed be assigned to Fe-rich defects on the
chemically ordered \feni{}(001) surface.

Mapping the tunneling current at a voltage just below the
low-energy edge of the resonance peak by plotting
$I(x,y,z_0,-0.4~V)$ from the complete I/V dataset provides a map
of the surface resonance intensity acquired in a distance of more
than 0.5 nm from the surface plane [Fig. 2(b)]. By evaluating the
similarity of the spectrum in each image point to the peaked
spectrum (A) and the ``flat'' spectrum (B) we find (see caption
Fig.~2) that the surface resonance peak is a minority property of
the surface, with the larger surface fraction showing no or only a
small peak, suggesting a correlation to the Fe-rich defects seen
in the atomically resolved images [Fig.~1(a)]. Direct comparison
of the classification by the tunneling spectra [contours in
Fig.~2(a,b)] with the constant current image [Fig. 2(a)]
demonstrates that the apparent height of the surface does not
obviously correlate with the map of the surface resonance
intensity. Although a weak correlation should exist, because the
composition of the surface monolayer contributes to the apparent
height as well, it is too weak to be relevant compared to the
height variation caused by the compounded effect of concentration
inhomogeneities in each of several monolayers below the surface
monolayer. Thus, the tunneling spectra provide a method to map the
chemical information independent of the geometric height
information on every image scale without the requirement to
resolve the surface atomically. We remark that, although
desirable, it would be extremely difficult to combine such a
spectroscopic analysis with the simultaneous acquisition of an
atomically resolved image, mainly due to the instability and short
lifetime of very sharp tips.

\section{DFT calculations}
\subsection{Surface relaxation} We first investigate the surface
geometry of the Fe-Ni alloy. The relation between crystal
structure and magnetic ordering is of crucial importance for
iron-based alloys. The calculations have been done using the
full-potential linearized augmented plane wave method \cite{wimm}
(FP-LAPW) code FLAIR \cite{FLAIR}.  Two approximations for the
density functional potential have been used: the local density
approximation (LDA) according to Vosko, Wilk, and Nusair\cite{VWN}
and the generalized gradient approximation (GGA) according to
Perdew, Burke and M. Ernzerhof \cite{pbe}. Spherical wave and
plane wave expansions were truncated at $l_{max} = 8$ and
$E_{cut}$ = 11.6~Ry respectively. The difference between the last
two charge density iterations was less than $ 10^{-6}$~ e~a.u.$^{
-3}$. In all the surface geometry optimizations, atomic positions
in the surface and two subsurface layers were fully
relaxed\cite{Ondra06}. The final relaxed structure was assumed
when the force on each atom was smaller than \mbox{5 meV/\AA }.
 Nine-layer ordered films have been used to simulate fcc(001) alloy
surfaces with different ordering. For all the calculations we have
used the experimental lattice constant $a$ = 3.58~\AA
~\cite{miodo79}. Minimizing the total energy as a function of
lattice constant for a NiFe$_3$ crystal yields the value $a$ =
3.60~\AA . Our model geometries are derived from a NiFe$_3$(001)
surface structure composed of $c(2\times 2)$ NiFe(001)-layers
alternating with pure Fe(001) layers. It should be noted here that
this corresponds to the Cu$_3$Au structure, also known as L1$_2$.
Our models are in accordance with STM experiments, which indicate
approximately the same amount of iron and nickel atoms in the top
surface layer. All model systems exhibit inversion symmetry and
include 18 atoms in the unit cell. Some models have only iron
atoms in the subsurface layer and differ in the geometry of deeper
layers, e.g. by interchanged Ni and Fe atom sites. We considered
also systems where the subsurface layer includes both type of
atoms.

Results of first principles atomic force minimizations for
different models\cite{Ondra06} can be summarized as follows: the
top surface NiFe ordered monolayer is buckled with Fe atoms pushed
outwards and Ni atoms pushed inwards. The buckling height in the
top surface layer, \mbox{0.07~to~0.12~\AA,} depends on the type of
atoms in subsurface layers. We may assume that also in the
disordered fcc(001) alloy surface the Fe atoms are pushed outwards
and their positions are determined  by the local chemical
composition of subsurface atoms. The atomically resolved STM
images essentially reflect the core positions of the Fe-Ni (001)
surface. The brighter spots can be associated with Fe atoms, the
darker with Ni atoms. We conclude that the subsurface composition
influences the top surface layer buckling as well as the first
interlayer distance $d_{12}$ [$\Delta d_{12} = d_{12} - d_{\perp
}^{bulk} \in (-0.07, +0.02)$~\AA ]\cite{Ondra06}. The presence of
Ni in the subsurface layer results in a decrease of the first
interlayer distance and also the buckling of the surface is
smaller. Different local chemical composition in the subsurface
explains observed brighter and darker nanometer-sized areas 
in the measured STM images.

Due to their reduced coordination, surface atoms in magnetic
systems  have higher magnetic moments as compared to the
bulk\cite{Free91}. It is well known that the surface magnetism
depends not only on the coordination number, but also on the
details of atomic arrangement. For our models, we have found that
the magnetic moments on the surface, 2.97~$\mu _B$ for Fe and 0.74
$\mu _B$ for Ni, are larger than the bulk values 2.61 $\mu _B$ and
0.68 $\mu _B$. The given values are local magnetic moments
obtained by integrating the charge distributions over the atomic
(muffin-tin) sphere. The calculated work function for the alloy is
4.50~eV.

\subsection{\label{surfelstruc}Surface electronic structure}
The peak in the tunneling conductivity found about 0.35~eV below
the Fermi level seems to indicate that a surface state or a
surface resonance exists on the \feni{} fcc(001) surface at this
energy. In order to investigate the surface electronic structure,
the spin-polarized local density of states (LDOS) has been
calculated for different surface geometries with FP-LAPW. We found
a surface resonance below the Fermi level, but its energy and
intensity was strongly dependent on the detailed film composition.
We needed thick film models (more than 30 layers) to achieve
well-converged results, rendering the computational demand for
FP-LAPW too high.

Therefore we have applied the tight-binding linear muffin-tin
orbital method (TB-LMTO) in the atomic sphere approximation (ASA)
to evaluate the surface alloy features. The TB-LMTO code is based
on the surface Green function formalism and incorporates realistic
boundary conditions.  The exchange-correlation potential due to
Vosko, Wilk, and Nusair \cite{VWN} has been employed. The dipole
barrier effects due to charges spilling out from the sample into
the vacuum are included in the formalism. The effect of randomness
in the sample is described in the framework of the 
CPA (coherent potential approximation), which is able to treat
arbitrary concentration profiles close to the sample surface. For
further details see Refs.~\onlinecite{Kudrnov93}
and~\onlinecite{book}.

The sample consists of two semi-infinite regions, the vacuum and
the substrate, sandwiching the intermediate region which also
includes few 'substrate' and 'vacuum' layers. The potentials in
the intermediate region are determined selfconsistently while for
the semi-infinite substrate bulk potentials are used. The vacuum
region is represented by a flat potential; and the substrate is a
binary alloy \feni{}. The intermediate region consists of 10
atomic layers, namely of three empty-sphere layers adjacent to the
vacuum region, two surface alloy layers in which the composition
can differ from that in bulk alloy, and five alloy layers in which
we assume Invar bulk composition. In this way, charge densities in
the intermediate region smoothly match those in the vacuum and
substrate regions. The so-called principal
layers\cite{Kudrnov93,book} consisted of two atomic layers. The
notation
$\text{Fe}_{x}\text{Ni}_{1-x}/\text{Fe}_{y}\text{Ni}_{1-y}/
\text{Fe}_{0.64}\text{Ni}_{0.36}$ will be used to describe the
composition of the two surface layers. We assume an ideal fcc
lattice with a lattice constant of $a=3.58$~\AA ~for all layers.
Thus, no lattice or layer relaxations are assumed. Changes of the
interatomic distances due to the surface relaxation are only
within few percent and do not affect the electronic structure of
the \feni{} alloy substantially. The buckling of Fe and Ni atoms
is also neglected. Possible errors introduced by this
approximation become small for Fe-rich alloy surfaces which are of
interest here.

The largest contribution to the tunneling current between the tip
and the sample comes from the regions around the point
$k_{\|}=\bar{\Gamma}$ of the surface Brillouin zone, where the
corresponding $k_{\|}$-resolved local density of electron states
at given energy tends to decay most slowly into the vacuum (cf.
eg. Ref. \onlinecite{FeSS}). Consequently, we start to analyze the
spin- and $k_{\|}$-resolved local density of states, also called
the Bloch spectral function (BSF), evaluated at the tip apex
position for $k_{\|}=\bar{\Gamma}$.

\begin{figure}[h]
\includegraphics[height=86mm]{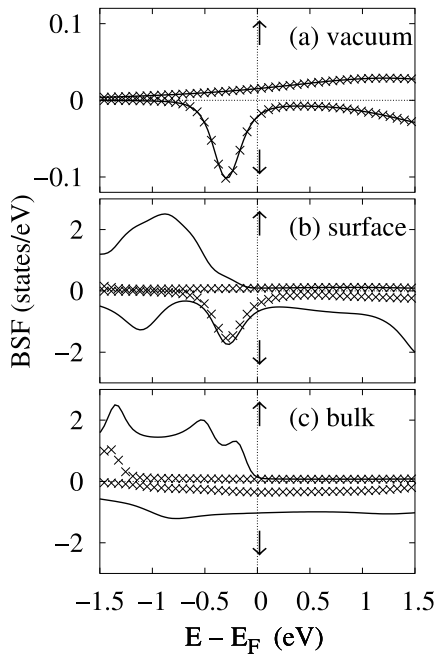}
\caption{\label{layerresolved} Solid lines: Layer- and
spin-resolved Bloch spectral functions at $\bar{\Gamma}$ for
$\text{Fe}/\text{Fe}_{0.64}\text{Ni}_{0.36}$ (a) in vacuum
3.6~\AA{} above the surface, (b) in the top surface layer, and (c)
in a layer deep in the bulk. Crosses denote the $A_1$-symmetry
component of the BSF. }
\end{figure}

The BSF for a pure iron surface layer on the \feni{} alloy is
shown in Fig.~\ref{layerresolved}. The results clearly demonstrate
that a surface resonance appears on iron-rich surfaces in an
energy range centered at -0.3~eV, in good agreement with the
experimental data. This surface resonance is formed by
minority-spin electron states in the central part of the surface
Brillouin zone and it has predominantly $d_{z^2}$ character. The
totally symmetric $A_1$-component of the BSF is also shown. This
$A_1$-component can be as well expressed as a sum of the $s$,
$p_z$, and $d_{z^2}$ orbital-resolved components. It is this
particular component that gives rise to the surface resonance. The
totally symmetric component is expected to decay into the vacuum
more slowly than the other components at the same energy do, and
this behavior is indeed observed. Let us note that the
$A_1$-component remains nonzero in the same energy range also
inside the bulk substrate, as shown in
Fig.~\ref{layerresolved}(c). The resonant electron states on the
surface can thus hybridize with corresponding bulk states.

\begin{figure}[h]
\includegraphics[height=86mm,angle=270]{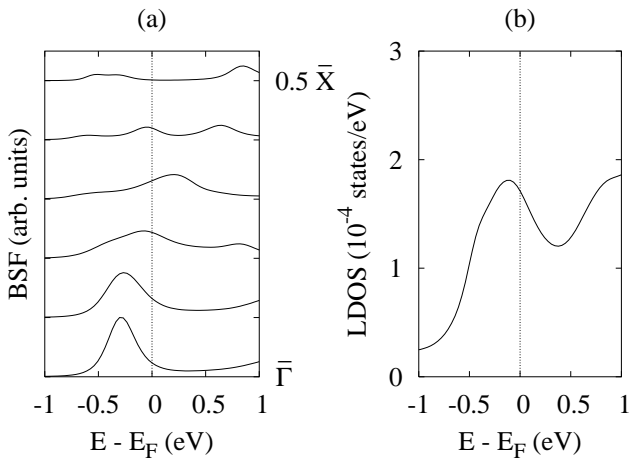}
\caption{\label{Fig4ab} (a) Minority-spin Bloch spectral functions
in vacuum 5.4~\AA{} above the
$\text{Fe}/\text{Fe}_{0.64}\text{Ni}_{0.36}$ surface.
The BSF is plotted for different $k_{\|}$-vectors
along the path from $k_{\|}=\bar{\Gamma}\equiv (0,0)$ (bottom) to
$k_{\|}=0.5\,\bar{\text{X}}\equiv (0.5,0)$ (top). The curves are
shifted vertically according to their corresponding
$k_{\|}$-vectors. (b) Local density of states in vacuum 5.4~\AA{}
above the $\text{Fe}/\text{Fe}_{0.64}\text{Ni}_{0.36}$ surface. }
\end{figure}

The energy dispersion of the surface resonance at the iron-covered
surface is illustrated in Fig.~\ref{Fig4ab}(a). The surface
resonance, well resolved at $k_{\|}=\bar{\Gamma}$, exhibits strong
dispersion along the $\bar{\Gamma} - \bar{X}$ direction in the
two-dimensional Brillouin zone. The LDOS is simply the
corresponding $k_{\|}$-integrated BSF and it is shown in
Fig.~\ref{Fig4ab}(b). The LDOS reproduces again a pronounced peak
for a low binding energy but its position is slightly shifted
upwards. This result follows from the fact that contributions from
the whole surface Brillouin zone are included in the LDOS on equal
footing. The sharp peak observed in the STS experiment indicates
that the tip apex acts in this case as a sharp $k_{\|}$=0-filter.
The contributions from the central part of the surface Brillouin
zone dominate. The broadening effects caused both by the energy
dispersion of the surface resonance and the tip - sample
interaction (not included in our analysis) are more important for
small tip - sample distances (below 5~\AA )\cite{FeSS}.

\begin{figure}
\includegraphics[height=86mm,angle=270]{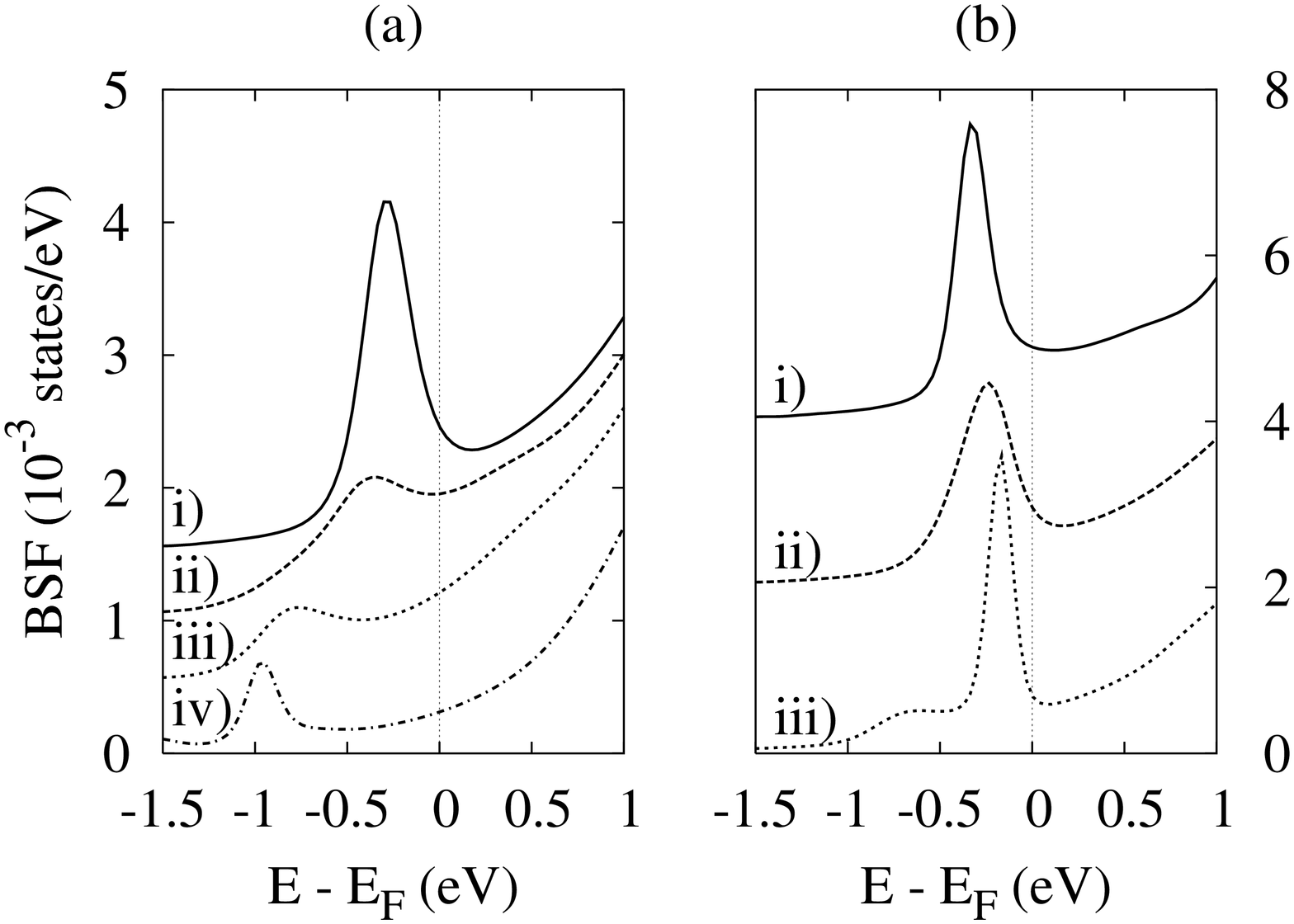}
\caption{\label{Fig5ab} Bloch spectral functions in vacuum
5.4~\AA{} above the surface. (a) The top surface layer composition
is changed: i) $\text{Fe}/\text{Fe}_{0.64}\text{Ni}_{0.36}$, ii)
$\text{Fe}_{0.75}\text{Ni}_{0.25}/\text{Fe}_{0.64}\text{Ni}_{0.36}$,
iii)
$\text{Fe}_{0.50}\text{Ni}_{0.50}/\text{Fe}_{0.64}\text{Ni}_{0.36}$,
and iv) $\text{Ni}/\text{Fe}_{0.64}\text{Ni}_{0.36}$.\\
(b) The subsurface layer composition is changed: i)
$\text{Fe}/\text{Fe}/\text{Fe}_{0.64}\text{Ni}_{0.36}$, ii)
$\text{Fe}/\text{Fe}_{0.50}\text{Ni}_{0.50}/\text{Fe}_{0.64}\text{Ni}_{0.36}$,
and iii) $\text{Fe}/\text{Ni}/\text{Fe}_{0.64}\text{Ni}_{0.36}$.
The curves are shifted vertically in order to be better
distinguishable.}
\end{figure}

The concentration dependence of the surface resonance is shown in
Fig.~\ref{Fig5ab}(a). The strength of the surface resonance is
reduced with increasing content of Ni-atoms in the topmost surface
layer: it almost disappears for less than 75\,\%~Fe in the surface
layer. Another surface resonance is formed on the Ni-rich
surfaces. This resonance is shifted to lower energies (about
$1$~eV below the Fermi level) and also corresponds to
$d_{z^2}$-like minority-spin states. We have also explored how the
surface resonance on the iron-rich surfaces depends on subsurface
composition of the sample [see Fig.~\ref{Fig5ab}(b)]. It turns out
that the energy position of the surface resonance depends only
weakly on the composition of the subsurface layer. We can conclude
that the position of the surface resonance in the spectra is
primarily determined by the composition of the topmost layer and,
thus, a good indicator for the local concentration in this layer.
This confirms the experimental observations.

\section{Conclusion}
The chemical contrast in atomically resolved STM images of fcc
Fe-Ni alloy surfaces can be explained by \textit{ab initio}
calculations which show that the (001) surface of \feni{} is
buckled: Fe atoms are shifted outwards; Ni atoms inwards. We find
that although the surface Fe concentration is $\approx{} $59\%,
the surface is mostly \c2x2{} ordered, which accommodates only
50\% Fe. The excess Fe atoms are inserted in frequent very short
Fe-rich anti-phase domain boundaries.

In contrast to surface alloys based on a homogeneous bulk,
however, the particular distribution of subsurface Fe and Ni atoms
additionally modulates the apparent height of the surface atoms in
the STM images. Compared to other alloys, this effect is rather
strong on the \feni{}(001) surface, suggesting significant
deviations from a random subsurface distribution, e.g., a
decomposition into regions with 50\% Fe and 100\% Fe similar to
the structure of the surface monolayer.

We have also observed a distinct peak in differential tunneling
conductance spectra at a negative sample bias of roughly 0.35~eV.
However, this peak is strong only in rather small regions on the
surface. On the basis of our \textit{ab initio} electronic
structure calculations, we attribute this peak to a
$d_{z^2}$-derived surface resonance band that is formed at the
$\bar{\Gamma}$-point of the surface Brillouin zone and in its
close vicinity with its lower edge 0.3~eV below the Fermi level.
The resonance occurs in model surfaces with high Fe content
(75\,\% and more in the topmost layer) and therefore indicates
regions on the surface which are Fe enriched, e.g., the mentioned
Fe-rich domain boundaries of the \c2x2{} ordered surface.

These results show that with certain limitations both surface and
subsurface chemical information can be extracted from STM and STS
images of the \feni{}(001) surfaces: (1) Constant current images
with atomic resolution at very low tunneling voltages provide
chemical contrast in an atom-by-atom fashion, which, however, is
mixed with a height modulation corresponding to the subsurface
composition. (2) Constant current images at high, in particular
positive, tunneling voltages, which are less sensitive to the
surface resonance intensity, map an apparent height, which is a
combined effect of the composition of the topmost two or even more
monolayers. (3) Current images at voltages just below the surface
resonance peak, which are very sensitive to the surface resonance
intensity, show the surface composition with a lateral resolution
on the nanometer scale without significant interference by the
subsurface composition.\\

\subsection*{Acknowledgments}

This work has been done within the project AVOZ10100520 of the ASCR. The
financial support provided by the Academy of Sciences of the Czech
Republic (Grant No. A1010214) and by the FWF (Austrian science fund) is
acknowledged.


\begin{thebibliography}{}

\bibitem{PtNi111} M. Schmid, H. Stadler, and P. Varga, Phys. Rev. Lett.
 {\bf 70}, 1441 (1993).

\bibitem{schmid02} M. Schmid and P. Varga, in {\em Alloy Surfaces and Surface Alloys}, ed. D. P. Woodruff, Elsevier, Amsterdam (2002), p.
118-151.

\bibitem{heben99}
E. L. D. Hebenstreit, W. Hebenstreit, M. Schmid, and P. Varga,
Surf. Sci {\bf 441}, 441 (1999).

\bibitem{PtRh} P.~T. Wouda, B.~E. Nieuwenhuys, M. Schmid, and P. Varga, Surf. Sci.
 {\bf 359}, 17 (1996).

\bibitem{cham92}
D. D. Chambliss and S. Chiang, Surf. Sci. {\bf 264}, L187 (1992).

\bibitem{nagl94}
C. Nagl, O. Haller, E. Platzgummer, M. Schmid, and P. Varga, Surf.
Sci. {\bf 321},  237 (1994).

\bibitem{niel03}
L. P. Nielsen, F. Besenbacher, I. Stensgaard, E. Laegsgaard, C.
Engdahl, P. Stoltze, K. W. Jacobsen, and J. K. Norskov, Phys. Rev.
Lett. {\bf 71}, 754 (1993).

\bibitem{cromm93}
M. F. Crommie, C. P. Lutz, and D. M. Eigler, Nature {\bf 363}, 524
(1993).

\bibitem{FeSS} J.~A. Stroscio, D.~T. Pierce, A. Davies, R.~J. Celotta, and
M.~Weinert,  Phys. Rev. Lett. {\bf 75}, 2960 (1995).

\bibitem{hofer01} W.~A. Hofer, J. Redinger, A. Biedermann, and P. Varga, Surf. Sci.
{\bf 482-485}, 1113 (2001).

\bibitem{CrFe} A. Davies, J.~A. Stroscio, D.~T. Pierce, and R.~J. Celotta,
  Phys. Rev. Lett. {\bf 76}, 4175 (1996).

\bibitem{Bieder96}
 A. Biedermann, O. Genser, W. Hebenstreit, M. Schmid, J. Redinger, R. Podloucky, and P.~Varga,
 Phys. Rev. Lett. {\bf 76}, 4179 (1996).

\bibitem{bisch01} M. M. J. Bischoff, C. Konvicka, A. J. Quinn, M. Schmid, J. Redinger,
 R. Podloucky, P. Varga, and H. van Kempen, Phys. Rev. Lett. {\bf 86}, 2396
 (2001); Surf. Sci. {\bf 513}, 9 (2002).


\bibitem{ters85}
J. Tersoff and D. R. Hamann, Phys. Rev. B {\bf 31}, 805 (1985).

\bibitem{post80}
M. Posternak, H. Krakauer, A.J. Freeman, and D.D. Koelling, Phys.
Rev. B, {\bf 21}, 5601 (1980).

\bibitem{rehb03}
A. Rehbein, D. Wegner, G. Kaindl, and A. Bauer, Phys. Rev. B {\bf
67}, 033403 (2003).

\bibitem{schi99}
Mark van Schilfgaarde, I. A.  Abrikosov, and B. Johansson, Nature
{\bf 400}, 46 (1999).

\bibitem{robe99}
J. L. Robertson, G. E. Ice, C. J. Sparks, X. Jiang, P. Zschack, F.
Bley, S. Lefebvre, and M. Bessiere, Phys. Rev. Lett. {\bf 82},
2911 (1999).

\bibitem{entel93}
P. Entel, E. Hoffmann, P. Mohn, K. Schwarz, and V. L. Moruzzi,
Phys. Rev. B {\bf 47}, 8706 (1993).

\bibitem{mohn91}
P. Mohn, K. Schwarz, and D. Wagner, Phys. Rev. B {\bf 43}, 3318
(1991).

\bibitem{willis05} R.~F. Willis and N. Janke-Gilman, Europhys. Lett. {\bf 69}, 411 (2005).

\bibitem{cris02} V. Crisan, P. Entel, H. Ebert, H. Akai, D.
D. Johnson, and J. B. Staunton, Phys. Rev. B {\bf 66}, 014416
(2002).

\bibitem{hank05}
T. H\"anke, M. Bode, S. Krause, L. Berbil-Bautista, and R.
Wiesendanger, Phys. Rev. B {\bf 72}, 085453 (2005).

\bibitem{crom93b}
M. F. Crommie, C. P. Lutz, and D. M. Eigler, Phys. Rev. B {\bf
48}, 2851 (1993).

\bibitem{wimm}
 M. Weinert, E. Wimmer, and A.~J. Freeman, Phys. Rev. B {\bf 26}, 4571 (1982);
     see also http://www.flapw.de.

\bibitem{FLAIR} URL:
http://www.uwm.edu/$\sim$weinert/flair.html.

\bibitem{VWN} S.~H. Vosko, L. Wilk, and M. Nusair, Can. J. Phys. {\bf 58}, 1200
(1980).

\bibitem{pbe} J.~P. Perdew, K. Burke, and M. Ernzerhof, Phys. Rev.
 Lett. {\bf 77}, 3865 (1996).

\bibitem{Ondra06}
 M. Ondr\'a\v{c}ek, F. M\'aca, J. Kudrnovsk\'y, and J. Redinger, Czech. J. Phys.
  {\bf 56}, 69 (2006).

\bibitem{miodo79} A.~P. Miodownik, J. Magn. Magn. Matter. {\bf 10},
126 (1979).

\bibitem{Free91}
A.~J. Freeman and R. Wu, J. Magn. Magn. Mater. {\bf 100}, 497
(1991).

\bibitem{Kudrnov93} J. Kudrnovsk\'{y}, I. Turek, V. Drchal, P. Weinberger,
S.~K. Bose, and A. Pasturel, Phys. Rev. B {\bf 47},  16525 (1993).

\bibitem{book}I. Turek, V. Drchal, J. Kudrnovsk\'y,  M. \v{S}ob,
and P.  Weinberger, {\em Electronic Structure of Disordered
Alloys, Surfaces and Interfaces}, Kluwer Academic Publishers,
Boston (1997).

\end{thebibliography}
\end{document}